\documentclass[sigconf,nonacm]{acmart}

\AtBeginDocument{%
  }

\usepackage{todonotes}
\usepackage{framed}
\usepackage[most]{tcolorbox}
\usepackage{xcolor}
\usepackage{soul}
\usepackage{geometry}
\usepackage{longtable}

\newcommand{\hlc}[2][yellow]{{%
    \colorlet{foo}{#1}%
    \sethlcolor{foo}\hl{#2}}%
}

\definecolor{themeRed}{HTML}{F25050}
\definecolor{themeBlue}{HTML}{506AF2}

\tcbset{
  highlights/.style={
    enhanced,
    colframe=blue!30,
    colback=blue!60,
    boxrule=0pt,
    corners=south,
    rounded corners=north,
    arc=1mm,
    boxsep=0pt,
    left=1mm,
    right=1mm,
    top=0.5mm,
    bottom=0.5mm, fontupper=\bfseries\color{white},
  },
}
\newtcolorbox{custombox}[1]{
	colback=gray!10,
	colframe=gray!70,
	left=1mm,
	right=1mm,
	top=1mm,
	bottom=1mm,
	fonttitle=\bfseries,
	arc=0mm,
	leftrule=1mm,
	rightrule=0mm,
	toprule=0mm,
	bottomrule=0mm,
	notitle,
	before=\par\smallskip\noindent,
	before upper={\textbf{#1: } },
}

\begin{document}


\title[Generating Synthetic Buggy Code Submissions for Computing Education]{LLM-itation is the Sincerest Form of Data: Generating Synthetic Buggy Code Submissions for Computing Education}


\author{Juho Leinonen}
\affiliation{%
  \institution{Aalto University}
  \city{Espoo}
  \country{Finland}}
\email{juho.2.leinonen@aalto.fi}
\orcid{0000-0001-6829-9449}

\author{Paul Denny}
\affiliation{%
  \institution{University of Auckland}
  \city{Auckland}
  \country{New Zealand}}
\email{paul@cs.auckland.ac.nz}
\orcid{0000-0002-5150-9806}

\author{Olli Kiljunen}
\affiliation{%
  \institution{Aalto University}
  \city{Espoo}
  \country{Finland}}
\email{olli.kiljunen@aalto.fi}
\orcid{0000-0002-1347-0699}

\author{Stephen MacNeil}
\affiliation{%
  \institution{Temple University}
  \city{Philadelphia}
  \state{PA}
  \country{USA}}
\email{stephen.macneil@temple.edu}
\orcid{0000-0003-2781-6619}

\author{Sami Sarsa}
\affiliation{%
  \institution{University of Jyväskylä}
  \city{Jyväskylä}
  \country{Finland}}
\email{sami.j.sarsa@jyu.fi}
\orcid{0000-0002-7277-9282}

\author{Arto Hellas}
\affiliation{%
  \institution{Aalto University}
  \city{Espoo}
  \country{Finland}}
\email{arto.hellas@aalto.fi}
\orcid{0000-0001-6502-209X}

\renewcommand{\shortauthors}{Leinonen et al.}

\begin{abstract}

There is a great need for data in computing education research. Data is needed to understand how students behave, to train models of student behavior to optimally support students, and to develop and validate new assessment tools and learning analytics techniques. However, relatively few computing education datasets are shared openly, often due to privacy regulations and issues in making sure the data is anonymous. Large language models (LLMs) offer a promising approach to create large-scale, privacy-preserving synthetic data, which can be used to explore various aspects of student learning, develop and test educational technologies, and support research in areas where collecting real student data may be challenging or impractical. This work explores generating synthetic buggy code submissions for introductory programming exercises using GPT-4o. We compare the distribution of test case failures between synthetic and real student data from two courses to analyze the accuracy of the synthetic data in mimicking real student data. Our findings suggest that LLMs can be used to generate synthetic incorrect submissions that are not significantly different from real student data with regard to test case failure distributions. Our research contributes to the development of reliable synthetic datasets for computing education research and teaching, potentially accelerating progress in the field while preserving student privacy.

\end{abstract}



\maketitle

\section{Introduction}

Computing education has witnessed a significant transformation with the rise of large language models (LLMs) \cite{denny2024computing}. LLMs have demonstrated remarkable capabilities in tasks relevant to computing educators and researchers, with a particular focus on their ability to solve introductory programming exercises \cite{Finnie-Ansley2022Robots}.  More recently, these capabilities have been demonstrated for more complex programming tasks \cite{finnie2023myai}, with current state-of-the-art models appearing able to solve almost all typical introductory programming exercises \cite{prather2023robots}.  This ability to generate correct code solutions is well-established, however less well explored is the potential of LLMs to deliberately create incorrect code.

Generating incorrect solutions is a relatively unexplored area but has many potential applications.  Incorrect code solutions can be used to create debugging exercises, which are known to be beneficial for learning \cite{whalley2021novice}. Additionally, they can help in generating synthetic datasets of student submissions, which include a mixture of correct and incorrect code. This is particularly valuable given the scarcity of openly shared programming education datasets, which are often constrained by strict privacy regulations and the challenges of de-identifying and anonymizing data \cite{leinonen2017preventing,edwards2023review}. Leveraging LLMs to create synthetic data can overcome these barriers, providing a new avenue for developing and validating educational tools and techniques without compromising student privacy.

Building on the extensive body of literature that has established LLMs' capability to generate correct solutions, in this work we explore the possibility of using these models to generate incorrect code submissions for introductory programming problems. We investigate various prompting strategies to determine which approaches produce submissions that most closely resemble real student data. Our evaluation focuses on the distribution of test case failures as a measure of similarity between synthetic and real data. This study encompasses two programming languages, C and Dart, and includes data from institutions across different countries.  The primary research question guiding this study is: 

\begin{quote}
    To what extent can generative AI models be used to generate synthetic incorrect code submissions for introductory programming exercises?
\end{quote}

We make two main contributions in this work.  First, we provide an analysis of the capability of LLMs to generate synthetic data for computing education research.  Second, we explore the effectiveness of prompting strategies in generating incorrect code submissions that mirror typical student errors.  This study contributes to the development of reliable synthetic datasets, which can facilitate research and teaching in computing education while preserving student privacy.

\section{Related Work}

\subsection{Bugs}

Programming errors, or bugs, constitute a well studied topic
in the computing education research literature.
Many studies have identified common mistakes made by novice programmers,
including, syntax errors, logic errors~\cite{ettles2018common},
or both~\cite{altadmri2015million}. Teaching students to debug has also gained a great deal of researchers' attention~\cite{mccauley2008debugging,li2019towards}.
Griffin's study \cite{griffin2019designing} exemplifies
that the use of intentionally erroneous code in instruction
is not, however, limited to teaching debugging but is suitable
to programming education more generally.

However, the research literature has, so far, only narrowly covered
how to simulate programming errors made by students.
Until the emergence of LLMs, perhaps the most promising approach
was automatic program mutation, provided by
mutation analysis tools that software engineers use in software testing.
Clegg et al. \cite{clegg2021empirical} found that mutant code has similar faults
as code written by students and is, thus,
a good aid when designing automated grading systems.
Perretta et al. \cite{perretta2022on}, on the other hand, used code mutation for
evaluating test suites written by students.
They also conclude that mutant code can simulate student code
to a reasonable extent.

\subsection{LLMs in Computing Education}

With the introduction of generative AI, educators can now produce high-quality, personalized learning materials at scale for their students. These models can produce diverse explanations~\cite{macneil2022generating} that students find engaging~\cite{macneil2023experiences} and that are often rated higher in quality compared to explanations generated by peers~\cite{leinonen2023comparing}. Students and instructors can also use the models to generate personally relevant analogies~\cite{bernstein2024like, bernstein2024analyzing}, programming assignments~\cite{sarsa2022automatic}, and to create multiple choice questions~\cite{doughty2024comparative, tran2023generating}. As such, generative AI has become a legitimate source of help for many computing students~\cite{hou2023effects, prather2023robots} with over 26\% of students using it on a daily basis~\cite{hou2023effects}. 

To further support students, researchers have also developed interactive systems to scaffold student's use of generative AI in classroom settings. This scaffolding is essential to prevent misuse~\cite{zastudil2023generative, lau2023from, becker2023programming} and to address challenges that some students face in effectively prompting and interpreting responses from generative AI~\cite{prather2024widening, hou2023effects}. Such systems include, for example, CodeAid~\cite{kazemitabaar2024codeaid}, CodeHelp~\cite{liffiton2023codehelp}, and Promptly~\cite{denny2024prompt}. Despite the emergence of tools that scaffold the use of generative AI, less research has been dedicated to investigating whether generative AI can be used to simulate student behaviors or generate synthetic student data. Markel et al. simulated student questions to train teaching assistants~\cite{markel2023gpteach}, highlighting a potential area for further investigation.

\begin{table*}[t]
\footnotesize
\caption{Summary of the C programming project functions from 2016 and 2017. Every function had 20 tests.}
\label{table:C_function_summary}
\begin{tabular}{c|p{7.5cm}|p{8.5cm}}
\toprule
\textbf{Year} & \textbf{Prototype Declaration} & \textbf{Description} \\
\midrule
2016 & \texttt{int PrimeBelow(int upper);} & Returns the largest prime number less than the given upper limit. \\
\midrule
2016 & \texttt{void Strikeout(char *hide, char *phrase);} & Modifies a phrase by striking out occurrences of the word specified in hide. \\
\midrule
2016 & \texttt{int KthLargest(int k, int *values, int numValues);} & Returns the k-th largest element from an array of integers. \\
\midrule
2016 & \texttt{Rectangle BoundingRectangle(Rectangle r1, Rectangle r2);} & Computes the smallest rectangle that can enclose two given rectangles. \\
\midrule
2016 & \texttt{int TallestVine(int seedA, int seedB, int days);} & Simulates the growth of vines over a specified number of days based on seed values. \\
\midrule
2017 & \texttt{double AverageSheep(int *counts);} & Calculates the average count of sheep over a period (similar to a ``rainfall'' computation). \\
\midrule
2017 & \texttt{int PrimeFactors(int n, int *factors);} & Determines the prime factors of a given integer, n. \\
\midrule
2017 & \texttt{void ConnectTwo(int maze[10][10]);} & Modifies a 10x10 2D array to show the shortest connection between two specified cells. \\
\midrule
2017 & \texttt{void DayTrader(int *prices, int len, int *run, int *runIndex);} & Identifies the best run of consecutive days for maximizing return on stock prices. \\
\midrule
2017 & \texttt{void AddOne(char *input, char *output);} & Increments an arbitrarily large numeric string storing the result in an output string. \\
\bottomrule
\end{tabular}
\end{table*}

\subsection{Generating Synthetic Data}

Synthetic data is highly useful for multiple data science related purposes, including releasing privacy-preserving data in sensitive domains, construction of datasets without unwanted biases present in real world data, and augmenting scarce real world data \cite{jordon2022synthetic}. Synthetic data generators have been studied in various fields such as finance \cite{assefa2020generating}, medicine \cite{goel2023llms}, and computer science~\cite{botta20123531,moller2024parrot}. Likewise, the usefulness of synthetic data generation has been noted in the fields of education and learning analytics \cite{berg2016role}, e.g., to evaluate knowledge tracing models \cite{piech2015deep,sarsa2022empirical}, to train performance prediction models \cite{dorodchi2019using,flanagan2022fine}, and to simulate student behaviour for further research \cite{moreno2023synthetic,zhan2023synthetic}. However, the focus in generating datasets has been on numerical or categorical data, and not textual data such as student submissions for open text or programming exercises.

Enter LLMs which excel in generating text resembling that of humans and can be easily prompted to produce specific kinds of texts. On their own, they already are highly versatile and capable synthetic data generators for textual data given the correct prompts~\cite{li2023synthetic,wagner2024power}.
As a prime example, in a recent study by M{\o}ller et al. \cite{moller2024parrot}, augmenting training data of classification models with synthetic data (using GPT-4 and LLama2) was found to outperform augmenting it with crowdsourced data on some NLP classification tasks, particularly on multi-class tasks or tasks with rare classes, and to be beneficial on others although not as much as crowdsourcing. 
They note that using LLMs directly for various tasks is mostly inferior to using an LLM that is fine-tuned (i.e., trained further) using synthesized data, a result echoed by Tang et al. \cite{tang2023does} who investigated the capabilities of LLMs for healthcare related tasks.

While it is straightforward to generate synthetic data with LLMs, generating data of high quality and variability can require specific prompting strategies or knowledge enhancement~\cite{long2024llms,park2024chatlang}. Evaluating the quality of generated open-ended textual data directly as opposed to, e.g., evaluating it through classification models, can be laborious, requiring manual evaluations or auxiliary models \cite{chung2023increasing,long2024llms}.

\section{Methods}

Our data on student test case failures is taken from two distinct contexts, at institutions in different countries teaching different programming languages.  This diverse data allows us to better understand how well our proposed synthetic data generation approach might generalize to new contexts. 

\subsection{Context and Data}

\subsubsection{C Context}

Our first set of data is obtained from a six-week C programming module which is part of an introductory programming course taught at the University of Auckland, a research university located in New Zealand.  The content of this six-week module focuses on fundamental topics including basic syntax, data types, operators, standard I/O, control structures, functions, arrays, strings and file I/O.  Students take part in weekly lab sessions, where they complete short programming exercises and receive immediate feedback from an auto-grader.  The module concludes with a programming project, contributing 12\% towards their final grade for the course, for which students do not receive feedback until after the submission deadline.  Our data for this research includes student submissions to this final project taken from two consecutive deliveries of the course in 2016 and 2017. The data used in this study comprises a total of 8598 submissions of which 2405 are incorrect (i.e., do not pass all the tests) from 1751 students. 

The programming project includes the requirement to implement five distinct functions of varying difficulty.  Students are encouraged to thoroughly test their code prior to submitting it for grading as code that does not compile is not graded, and credit is only awarded for functions which pass all 20 of the tests in the corresponding test suite.  Table \ref{table:C_function_summary} lists the prototype declarations, along with brief descriptors (students and the LLM were provided more detailed specifications), for the ten functions (five from each year) for which we analyze student submissions and test case failures.

\subsubsection{Dart Context}

The second dataset comes from an online course platform that hosts a variety of courses offered by Aalto University, a research university located in Finland. For this study, ten exercises from two different courses were chosen. Both courses use the Dart programming language. The first course is an open online introductory programming course where participants are typically novices without any prior programming experience. The course teaches students basic programming concepts such as variables, printing output and reading input, conditional statements, iteration, lists, and functions. The course is worth 2 ECTS credits which corresponds to about 50-60 hours of workload. 

The second course is an advanced programming course where students are expected to know basic programming in some programming language. The goal of the second course is to teach students about developing software that supports a wide variety of devices using Dart and Flutter. The course is worth 5 ECTS credits which corresponds to about 135 hours of workload. As the participants are not expected to know Dart or Flutter, the course has a short introduction to key parts of the Dart programming language.

The introductory course is taught in Finnish while the advanced course is taught in English. For this article, we have translated the problem names and descriptions into English; however, the original language was used when prompting the LLM.

Both courses can be completed fully online and contain multiple small programming exercises embedded within the online materials. The data used for this study comes from a sample of ten of these small exercises. Table~\ref{tab:Dart_exercises} shows brief descriptions of the exercises used in this study. Students and the LLM were provided the actual, more comprehensive problem descriptions. The data comprises a total of 44742 submissions of which 24719 are incorrect (i.e., do not pass all the tests) from 5322 students (5063 from the introductory course and 259 from the advanced course).

\begin{table*}[t]
\footnotesize
\caption{Summary of Dart Programming Exercises.}
\label{tab:Dart_exercises}
\begin{tabular}{l|p{2.5cm}|p{10cm}|c}
\toprule
\textbf{Course} & \textbf{Exercise} & \textbf{Description} & \textbf{\# of tests} \\
\midrule
Introductory & Grade as text & Ask the user for a numerical grade and print the corresponding textual description of the grade. & 6\\
\midrule
Introductory & Sum of three numbers & Ask the user for three numbers and then print the sum of the numbers. & 2 \\
\midrule
Introductory & Ask for password & Write a function that takes a correct password as a parameter and then asks the user for the password until they input the correct password. & 3 \\
\midrule
Introductory & Sum of positive numbers & Write a function that takes in a list and returns the sum of the positive numbers in the list. & 2 \\
\midrule
Introductory & Authentication & Ask the user for specified username and password, and print different messages to the user depending on whether they input the correct username (``admin'') and/or password (``radish''). & 3 \\
\midrule
Advanced & Average of positives & Return the average value of positive numbers in a given list (similar to the Rainfall problem). & 4 \\
\midrule
Advanced & Budget check & Given two doubles, budget and spending, print whether the budget is okay or not. & 3 \\
\midrule
Advanced & Mystery function & Write a function that returns a string depending on which number is passed to the function and whether that number is divisible by 5 or 7 or both (similar to the FizzBuzz problem). & 5 \\
\midrule
Advanced & Sum with formula & Write a function that takes in two numbers and returns the written sum formula of those two numbers (e.g., for input 1 and 2, returns ``1+2=3''). & 2 \\
\midrule
Advanced & Video and playlist & Implement two classes, Video and Playlist. A video has a name (String) and duration in seconds (int) and a toString method. A playlist contains a list of videos and has methods for adding videos, checking if a video is on the playlist, and for returning the total duration of the playlist. & 3 \\
\hline
\end{tabular}
\end{table*}

\subsection{Prompting the LLM}

\begin{figure}[t!]
\centering
\begin{framed}
\raggedright
Problem description:\\
\textbf{<Problem description>}
\vspace{\baselineskip}
\\
\hlc[cyan]{Test cases:}\\
\hlc[cyan]{\textbf{<List of test cases>}}
\vspace{\baselineskip}
\\
\hlc[yellow]{Test case failure frequencies:}\\
\hlc[yellow]{\textbf{<List of failure frequencies for each test case>}}
\vspace{\baselineskip}
\\
Your task:\\
Please generate five incorrect solutions to this programming problem that include one or more semantic bugs. Place the delimiter CODE\_START before every solution example you'll generate and CODE\_END at the end of the solution code to help me extract just the generated code. Importantly, it should be possible to compile the incorrect solutions and it should be possible to run unit tests for the code. \hlc[yellow]{When generating the solutions, please try to follow the distribution of failing tests given above under ``Test case failure frequencies''.} Use the \textbf{<Dart/C>} programming language.
\end{framed}
\caption{The prompts used in the study. The baseline prompt did not include the blue and yellow highlighted parts. The test-case-informed prompt included the blue highlighted part on top of the baseline prompt. The frequency-informed prompt included both the blue and yellow highlighted parts on top of the baseline prompt. The bolded parts indicate variables for which content depended on the exercise.}
\label{fig:prompts}
\end{figure}

For this study, we chose to use the GPT-4o large language model\footnote{More specifically, the GPT-4o-2024-05-13 version.}, which at the time of writing, was the state-of-the-art model according to online leaderboards\footnote{According to the LMSYS Chatbot Arena Leaderboard, accessed July 20th, 2024: \url{https://chat.lmsys.org/?leaderboard}}. We evaluate three different prompts to explore how prompt engineering affects the generated incorrect solutions. The prompts used in this study are the following (for the exact phrasing, see Figure~\ref{fig:prompts}).

\begin{itemize}
    \item A \textit{baseline} prompt that just asks the model to generate semantically incorrect solutions to the problem.
    \item A \textit{test-case-informed} prompt where the model is also provided the test cases for the exercise.
    \item A \textit{frequency-informed} prompt where the model is provided both the test cases and the frequencies of how often incorrect submissions fail each specific test case. In addition, the model is instructed to try follow the distribution of the failure frequencies.
\end{itemize}

In our study, we only focus on semantic bugs. There are a few reasons for this. Firstly, we believe that semantic bugs are more interesting for potential debugging exercises. Secondly, if the code is not syntactically correct, then it would not be possible to run the test suite against the generated code, making it infeasible to evaluate the distribution of test case failures for the generated synthetic data. This is also why for all prompts, we explicitly tell the model that the generated solutions should compile and that it should be possible to run unit tests for the generated code.

When generating the submissions, for each combination of prompt, exercise, and context, we generate five batches of five incorrect solutions (i.e., 25 total for each unique combination of prompt, exercise and context). This results in a total of 3 prompts $\times$ 5 batches $\times$ 5 solutions $\times$ 10 exercises $\times$ 2 contexts = 1500 generated solutions.

\begin{table*}[t!]
    \centering
        \caption{Results of the analysis. The ``Real'' column shows statistics for the real student data (only incorrect solutions). The other three columns show the statistics for each of the three prompts we used. For each exercise, there were multiple unit tests. The range, mean, and standard deviation of the percentage of buggy solutions that pass at least one unit test are shown for each condition (real, baseline, test-case informed, frequency-informed). In addition, the average differences (deltas) between the real data and the synthetic data are shown for the means and standard deviations.}
    \label{tab:results}
    \begin{tabular}{ll|rrr|rrr|rrr|rrr}
\toprule
& & \multicolumn{3}{c|}{Real} & \multicolumn{3}{c|}{Baseline} & \multicolumn{3}{c|}{Test-case-informed} & \multicolumn{3}{c}{Frequency-informed} \\
Language & Exercise & \multicolumn{1}{c}{Range} & \multicolumn{1}{c}{$\mu$} & \multicolumn{1}{c|}{$\sigma$} & \multicolumn{1}{c}{Range} & \multicolumn{1}{c}{$\mu$} & \multicolumn{1}{c|}{$\sigma$} & \multicolumn{1}{c}{Range} & \multicolumn{1}{c}{$\mu$} & \multicolumn{1}{c|}{$\sigma$} & \multicolumn{1}{c}{Range} & \multicolumn{1}{c}{$\mu$} & \multicolumn{1}{c}{$\sigma$} \\
\midrule
C & Prime Below & [50, 92] & 81.1 & 12.6 & [58, 100] & 80.5 & 9.9 & [50, 100] & 68.8 & 14.7 & [61, 100] & 83.7 & 9.9 \\
C & Strikeout & [29, 75] & 57.5 & 13.1 & [0, 96] & 10.2 & 29.4 & [0, 83] & 9.5 & 25.2 & [0, 96] & 10.0 & 29.4 \\
C & Kth Largest & [43, 73] & 61.2 & 8.7 & [40, 96] & 66.2 & 19.0 & [48, 83] & 62.3 & 12.5 & [56, 96] & 70.0 & 14.8 \\
C & Bounding Rectangle & [22, 74] & 52.7 & 12.2 & [0, 28] & 8.4 & 7.6 & [4, 32] & 12.6 & 7.5 & [12, 44] & 22.4 & 9.7 \\
C & Tallest Vine & [31, 49] & 38.3 & 5.7 & [4, 62] & 40.2 & 24.0 & [0, 64] & 40.4 & 26.1 & [4, 76] & 50.2 & 29.7 \\
C & Average Sheep & [15, 84] & 70.8 & 17.4 & [32, 77] & 49.7 & 15.7 & [35, 83] & 53.5 & 18.0 & [42, 88] & 61.4 & 18.4 \\
C & Prime Factors & [24, 79] & 61.7 & 18.3 & [61, 74] & 69.4 & 4.7 & [39, 87] & 61.2 & 12.6 & [52, 81] & 66.7 & 9.4 \\
C & Connect Two & [23, 59] & 35.7 & 10.2 & [20, 30] & 23.5 & 3.3 & [23, 41] & 27.3 & 5.7 & [6, 24] & 8.7 & 5.0 \\
C & Day Trader & [30, 74] & 54.9 & 17.9 & [8, 28] & 11.8 & 5.3 & [0, 12] & 2.2 & 4.0 & [16, 28] & 19.8 & 4.2 \\
C & Add One & [23, 49] & 40.0 & 8.9 & [24, 76] & 50.0 & 19.9 & [76, 80] & 77.6 & 2.0 & [44, 96] & 73.0 & 22.4 \\
\midrule
\multicolumn{5}{l|}{Average deltas to real data for mean and standard deviation.} & & \multicolumn{1}{r}{19.3} & \multicolumn{1}{r|}{9.8} & & \multicolumn{1}{r}{22.0} & \multicolumn{1}{r|}{7.5} & & \multicolumn{1}{r}{21.1} & \multicolumn{1}{r}{9.4} \\
\midrule
Dart & Grade as text & [20, 48] & 38.7 & 9.2 & [44, 72] & 62.0 & 9.2 & [44, 76] & 58.0 & 11.3 & [44, 72] & 64.7 & 9.9 \\
Dart & Average of positives & [31, 46] & 38.0 & 6.6 & [24, 64] & 39.0 & 15.1 & [24, 48] & 33.0 & 9.9 & [8, 64] & 21.0 & 16.3 \\
Dart & Budget check & [24, 38] & 30.0 & 5.9 & [12, 40] & 21.3 & 13.2 & [12, 44] & 30.7 & 13.6 & [32, 64] & 49.3 & 13.2 \\
Dart & Sum of three numbers & [1, 2] & 1.5 & 0.5 & [0, 0] & 0.0 & 0.0 & [0, 0] & 0.0 & 0.0 & [4, 4] & 4.0 & 0.0 \\
Dart & Ask for password & [2, 31] & 17.3 & 11.9 & [0, 0] & 0.0 & 0.0 & [0, 0] & 0.0 & 0.0 & [0, 0] & 0.0 & 0.0 \\
Dart & Mystery function & [0, 82] & 64.2 & 32.1 & [4, 80] & 48.0 & 27.8 & [8, 68] & 44.0 & 20.2 & [8, 52] & 41.6 & 16.9 \\
Dart & Sum of positive numbers & [1, 6] & 3.5 & 2.5 & [24, 24] & 24.0 & 0.0 & [28, 32] & 30.0 & 2.0 & [20, 28] & 24.0 & 4.0 \\
Dart & Sum with formula & [0, 0] & 0.0 & 0.0 & [0, 4] & 2.0 & 2.0 & [0, 0] & 0.0 & 0.0 & [0, 0] & 0.0 & 0.0 \\
Dart & Authentication & [24, 58] & 38.0 & 14.5 & [24, 44] & 33.3 & 8.2 & [28, 44] & 34.7 & 6.8 & [32, 72] & 49.3 & 16.8 \\
Dart & Video and playlist & [33, 71] & 50.7 & 15.6 & [16, 40] & 24.0 & 11.3 & [24, 44] & 34.7 & 8.2 & [40, 52] & 45.3 & 5.0 \\
\midrule
\multicolumn{5}{l|}{Average deltas to real data for mean and standard deviation.} & & \multicolumn{1}{r}{12.2} & \multicolumn{1}{r|}{4.8} & & \multicolumn{1}{r}{11.0} & \multicolumn{1}{r|}{5.3} & & \multicolumn{1}{r}{14.2} & \multicolumn{1}{r}{6.0} \\
\bottomrule
\end{tabular}
\end{table*}

\subsection{Analysis}

We only include incorrect submissions in our analysis. This is done as our focus in this work is to generate incorrect synthetic solutions. For all the generated synthetic submissions, as well as for the real data used as a comparison point, we ran the unit test suites that had been used for those exercises when they were part of the course. For each individual unit test, we then calculated the percentage of cases where the solutions pass and fail the unit test, i.e. a unit test ``pass rate''. This gives us one percentage, or pass rate, for each unit test. As there are hundreds of unit tests altogether for the 20 exercises analyzed in this work, we calculate the minimum, maximum, mean, and standard deviation of these unit test pass rates for each exercise (which each comprise multiple unit tests). This way, we can compare if the unit test pass rates between the real and the synthetic data are different with regard to the pass rate range (i.e., minimum and maximum), mean, and standard deviation.

Even though we asked the model to produce code that unit tests could be run against, the model would sometimes produce code that crashed. For C, there were 48 (out of 750) generated programs that crashed when trying to run the test suite (typically due to a segmentation fault). These were ignored in calculating the statistics. 

To analyze the difference between the generated synthetic data and real data statistically, we conduct a Kruskal-Wallis H test between the test pass rates between all four distributions (real, baseline, test-case-informed, and frequency-informed) for both programming languages separately. In case either of the Kruskal-Wallis H test results suggests that the distributions are statistically significantly different using an alpha threshold of 0.05, we conduct pairwise Mann-Whitney U tests between the real data and each synthetic dataset separately to analyze which of the synthetic datasets are significantly different from the real data. As we do multiple statistical comparisons, we employ the Bonferroni correction to avoid finding spurious statistically significant differences.

\section{Results and Discussion}

The results of the analysis are shown in Table~\ref{tab:results}. Many interesting observations can be made based on the table. First, there seem to be differences between exercises in how well the model can generate incorrect solutions to the exercise. Some exercises seem hard for the model to solve ``partially incorrectly'', i.e., to generate a bug that allows some tests to pass. This is the case, for example, for the ``Ask for password'' Dart exercise and the ``Day Trader'' C exercise. For the former, the mean pass rate in the real data is 17.3\%, but all the LLM-generated incorrect solutions always fail all the tests. For the latter, the mean pass rate in the real data is 54.9\%, which is considerably higher than the mean pass rate for all three prompts: 11.8\%, 2.2\%, and 19.8\% for the baseline, test-case informed, and frequency-informed prompts respectively. This suggests that for these two exercises, the bugs generated by the model tend to cause most of the tests to fail, while in the real data, student bugs are more subtle and only cause part of the tests to fail. This finding is similar to synthesized code that is aimed to be correct where it has also been found that LLM performance is problem dependent~\cite{liu2024your}.

When considering the results, the number of test cases should be taken into account for the Dart data (all the C exercises had exactly 20 test cases each). For example, the ``Sum of three numbers'' and the ``Sum with formula'' exercises both only had two test cases. For both of these exercises, the tests mainly check that the student has not hard coded the response, and thus most bugs (other than hard coding) will cause both tests to fail concurrently. For these two exercises, the very low ranges and means of unit test pass rates suggest that real buggy submissions almost exclusively fail both tests, i.e., it is very rare that one test would pass and the other not.

Somewhat surprisingly, there do not seem to be large differences between the different prompts in how well they work for generating synthetic incorrect submissions. This is most visible by looking at the average deltas between the real data and the synthetic data. This is confirmed for the Dart data by a Kruskal-Wallis H test (H = 0.87, p = 0.83), which suggests that all four distributions are statistically equivalent. However, for the C data, the results of the Kruskal-Wallis H test (H = 28.4, p < 0.0001) suggest that at least one distribution is significantly different from the others. Pairwise Mann-Whitney U tests between the real data and each synthetic dataset separately reveal that both the baseline (U = 25739.0, p < 0.0001) and the test-case-informed (U = 25036.5, p < 0.0001) synthetic datasets are significantly different from the real data. However, the difference is not significant for the frequency-informed synthetic dataset with our threshold for significance alpha = 0.05 (U = 22816.0, p = 0.07). For our study, these results imply that all three prompts led to ``good'' synthetic data for Dart (as it was not significantly different from real data), while only the frequency-informed prompt led to ``good'' synthetic data for C.

As providing test case information and failure distribution to GPT-4o does not always appear to help it generate submissions with more similar distributions to real data, more sophisticated prompt engineering approaches could be a useful area to explore, at least for the current generation of state-of-the-art models. Previous work has found differences between student and LLM-generated code, for example, in what constructs and keywords are used~\cite{hoq2024detecting,denny2023can}. In general, generating synthetic content using LLMs risks monotonicity, especially if content is generated without controls aimed at increasing the diversity of the generated content~\cite{long2024llms}.

The finding that the synthetic Dart data seems to be more similar to the real student data with regard to test case failure distributions is corroborated by looking at the average deltas. For the Dart data, the average deltas between the mean pass rates for the real data and the synthetic data are considerably lower (12.2\% for baseline, 11.0\% for test-case-informed, and 14.2\% for frequency-informed) than for the C data (19.3\% for baseline, 22.0\% for test-case-informed, and 21.1\% for frequency-informed). This finding is surprising as the model has likely been trained with more C code than Dart code since C is a vastly more common programming language compared to Dart. This suggests that it might be harder for the model to generate semantic bugs that are similar to bugs found in student programs for C than for Dart. On the other hand, the Dart exercises are less complex than the C exercises, which could also contribute to the observation.

Two of the C exercises had diagrams in their problem descriptions that were not shown to the model during prompting. For the ``Tallest Vine'' exercise, this does not seem to have been a problem for the model as the mean pass rates for the synthetic data are higher than for the real data (40.2\% for baseline, 40.4\% for test-case-informed, and 50.2\% for frequency-informed versus 38.3\% for real data), suggesting the model was able to generate solutions that pass some of the tests. However, for the ``Bounding Rectangle'' exercise, this might explain why the mean pass rates for the synthetic data are considerably lower than for the real data (8.4\% for baseline, 12.6\% for test-case-informed, and 22.4\% for frequency-informed versus 52.7\% for real data).

\section{Limitations}

There are some limitations to this work. We only ask the model to generate incorrect solutions. In both contexts, a large portion of submissions pass all the tests (45\% of submissions for Dart and 72\% of submissions for C). Our analysis does not look at whether the model can generate realistic correct solutions, which is left for future research. Prior work suggests that LLMs can solve most introductory programming exercises correctly~\cite{prather2023robots}, although LLM-generated solutions have distinct patterns that make it possible to distinguish them from student-generated solutions~\cite{hoq2024detecting}. Thus, future work should study whether LLMs can be used to generate realistic synthetic correct solutions.

We only evaluate the similarity of the synthetic data to the real data with regard to test case failure distributions. For example, we do not look at constructs used in the code, what strategies are employed in the program to solve the problem, or the actual bugs in the code.

Some of the test suites of the Dart exercises were not very comprehensive, only including a couple of tests. This means that some bugs that the LLMs generate might not be captured by the test suite. For the C exercises, two of them had diagrams in the problem descriptions that were not shown to the LLM. Thus, the LLM was not provided the same information as the students, which could have made it more difficult for the LLM to generate the incorrect solutions, potentially affecting the results.

\section{Conclusions}

We investigated the capability of generative AI models in generating synthetic incorrect code submissions to programming exercises. This could be useful for creating debugging exercises for students and for generating synthetic datasets for research purposes. Our findings suggest that LLMs can be used to generate synthetic incorrect submissions that are not significantly different from real student data with regard to test case failure distributions. This provides evidence that LLMs could be used for generating satisfiably diverse synthetic code submission data, potentially lowering barriers to conducting research with such data, and making it easier to provide students with debugging practice.

However, more research is necessary to explore the closeness of LLM generated synthetic code submissions to that of real student data in more detail, such as what in the code makes the test cases fail and can the patterns in synthetic code submissions be made to more closely resemble that of real student code submissions.

\begin{acks}
This research was supported by the Research Council of Finland (Academy Research Fellow grant number 356114).
\end{acks}

\balance
\bibliographystyle{ACM-Reference-Format}
\bibliography{refs}


\begin{thebibliography}{52}


\ifx \showCODEN    \undefined \def \showCODEN     #1{\unskip}     \fi
\ifx \showDOI      \undefined \def \showDOI       #1{#1}\fi
\ifx \showISBNx    \undefined \def \showISBNx     #1{\unskip}     \fi
\ifx \showISBNxiii \undefined \def \showISBNxiii  #1{\unskip}     \fi
\ifx \showISSN     \undefined \def \showISSN      #1{\unskip}     \fi
\ifx \showLCCN     \undefined \def \showLCCN      #1{\unskip}     \fi
\ifx \shownote     \undefined \def \shownote      #1{#1}          \fi
\ifx \showarticletitle \undefined \def \showarticletitle #1{#1}   \fi
\ifx \showURL      \undefined \def \showURL       {\relax}        \fi
\providecommand\bibfield[2]{#2}
\providecommand\bibinfo[2]{#2}
\providecommand\natexlab[1]{#1}
\providecommand\showeprint[2][]{arXiv:#2}

\bibitem[Altadmri and Brown(2015)]%
        {altadmri2015million}
\bibfield{author}{\bibinfo{person}{Amjad Altadmri} {and} \bibinfo{person}{Neil~CC Brown}.} \bibinfo{year}{2015}\natexlab{}.
\newblock \showarticletitle{37 million compilations: Investigating novice programming mistakes in large-scale student data}. In \bibinfo{booktitle}{\emph{Proc. of the 46th ACM Technical Symp. on Computer Science Education}}. \bibinfo{pages}{522--527}.
\newblock


\bibitem[Assefa et~al\mbox{.}(2020)]%
        {assefa2020generating}
\bibfield{author}{\bibinfo{person}{Samuel~A Assefa}, \bibinfo{person}{Danial Dervovic}, \bibinfo{person}{Mahmoud Mahfouz}, \bibinfo{person}{Robert~E Tillman}, \bibinfo{person}{Prashant Reddy}, {and} \bibinfo{person}{Manuela Veloso}.} \bibinfo{year}{2020}\natexlab{}.
\newblock \showarticletitle{Generating synthetic data in finance: opportunities, challenges and pitfalls}. In \bibinfo{booktitle}{\emph{Proceedings of the First ACM International Conference on AI in Finance}}. \bibinfo{pages}{1--8}.
\newblock


\bibitem[Becker et~al\mbox{.}(2023)]%
        {becker2023programming}
\bibfield{author}{\bibinfo{person}{Brett~A Becker}, \bibinfo{person}{Paul Denny}, \bibinfo{person}{James Finnie-Ansley}, \bibinfo{person}{Andrew Luxton-Reilly}, \bibinfo{person}{James Prather}, {and} \bibinfo{person}{Eddie~Antonio Santos}.} \bibinfo{year}{2023}\natexlab{}.
\newblock \showarticletitle{Programming Is Hard -- Or at Least It Used to Be: Educational Opportunities And Challenges of AI Code Generation}. In \bibinfo{booktitle}{\emph{Proc. of the 54th ACM Technical Symp. on Computer Science Education V. 1}}.
\newblock


\bibitem[Berg et~al\mbox{.}(2016)]%
        {berg2016role}
\bibfield{author}{\bibinfo{person}{Alan~Mark Berg}, \bibinfo{person}{Stefan~T Mol}, \bibinfo{person}{G{\'a}bor Kismih{\'o}k}, {and} \bibinfo{person}{Niall Sclater}.} \bibinfo{year}{2016}\natexlab{}.
\newblock \showarticletitle{The role of a reference synthetic data generator within the field of learning analytics.}
\newblock \bibinfo{journal}{\emph{Journal of Learning Analytics}} \bibinfo{volume}{3}, \bibinfo{number}{1} (\bibinfo{year}{2016}), \bibinfo{pages}{107--128}.
\newblock


\bibitem[Bernstein et~al\mbox{.}(2024a)]%
        {bernstein2024like}
\bibfield{author}{\bibinfo{person}{Seth Bernstein}, \bibinfo{person}{Paul Denny}, \bibinfo{person}{Juho Leinonen}, \bibinfo{person}{Lauren Kan}, \bibinfo{person}{Arto Hellas}, \bibinfo{person}{Matt Littlefield}, \bibinfo{person}{Sami Sarsa}, {and} \bibinfo{person}{Stephen MacNeil}.} \bibinfo{year}{2024}\natexlab{a}.
\newblock \showarticletitle{"Like a Nesting Doll": Analyzing Recursion Analogies Generated by CS Students Using Large Language Models}. In \bibinfo{booktitle}{\emph{Proc. of the 2024 on Innovation and Technology in CS Education V. 1}}. \bibinfo{pages}{122--128}.
\newblock


\bibitem[Bernstein et~al\mbox{.}(2024b)]%
        {bernstein2024analyzing}
\bibfield{author}{\bibinfo{person}{Seth Bernstein}, \bibinfo{person}{Paul Denny}, \bibinfo{person}{Juho Leinonen}, \bibinfo{person}{Matt Littlefield}, \bibinfo{person}{Arto Hellas}, {and} \bibinfo{person}{Stephen MacNeil}.} \bibinfo{year}{2024}\natexlab{b}.
\newblock \showarticletitle{Analyzing Students' Preferences for LLM-Generated Analogies}. In \bibinfo{booktitle}{\emph{Proc. of the 2024 on Innovation and Technology in Computer Science Education V. 2}}.
\newblock


\bibitem[Botta et~al\mbox{.}(2012)]%
        {botta20123531}
\bibfield{author}{\bibinfo{person}{Alessio Botta}, \bibinfo{person}{Alberto Dainotti}, {and} \bibinfo{person}{Antonio Pescapé}.} \bibinfo{year}{2012}\natexlab{}.
\newblock \showarticletitle{A tool for the generation of realistic network workload for emerging networking scenarios}.
\newblock \bibinfo{journal}{\emph{Computer Networks}} \bibinfo{volume}{56}, \bibinfo{number}{15} (\bibinfo{year}{2012}), \bibinfo{pages}{3531--3547}.
\newblock
\showISSN{1389-1286}


\bibitem[Chung et~al\mbox{.}(2023)]%
        {chung2023increasing}
\bibfield{author}{\bibinfo{person}{John Chung}, \bibinfo{person}{Ece Kamar}, {and} \bibinfo{person}{Saleema Amershi}.} \bibinfo{year}{2023}\natexlab{}.
\newblock \showarticletitle{Increasing Diversity While Maintaining Accuracy: Text Data Generation with Large Language Models and Human Interventions}. In \bibinfo{booktitle}{\emph{Proceedings of the 61st Annual Meeting of the Association for Computational Linguistics (Volume 1: Long Papers)}}. \bibinfo{pages}{575--593}.
\newblock


\bibitem[Clegg et~al\mbox{.}(2021)]%
        {clegg2021empirical}
\bibfield{author}{\bibinfo{person}{Benjamin~Simon Clegg}, \bibinfo{person}{Phil McMinn}, {and} \bibinfo{person}{Gordon Fraser}.} \bibinfo{year}{2021}\natexlab{}.
\newblock \showarticletitle{An Empirical Study to Determine if Mutants Can Effectively Simulate Students' Programming Mistakes to Increase Tutors' Confidence in Autograding}. In \bibinfo{booktitle}{\emph{Proc. of the 52nd ACM Technical Symp. on Computer Science Education}}. \bibinfo{pages}{1055–1061}.
\newblock


\bibitem[Denny et~al\mbox{.}(2023)]%
        {denny2023can}
\bibfield{author}{\bibinfo{person}{Paul Denny}, \bibinfo{person}{Hassan Khosravi}, \bibinfo{person}{Arto Hellas}, \bibinfo{person}{Juho Leinonen}, {and} \bibinfo{person}{Sami Sarsa}.} \bibinfo{year}{2023}\natexlab{}.
\newblock \showarticletitle{Can we trust AI-generated educational content? comparative analysis of human and AI-generated learning resources}.
\newblock \bibinfo{journal}{\emph{arXiv preprint arXiv:2306.10509}} (\bibinfo{year}{2023}).
\newblock


\bibitem[Denny et~al\mbox{.}(2024a)]%
        {denny2024prompt}
\bibfield{author}{\bibinfo{person}{Paul Denny}, \bibinfo{person}{Juho Leinonen}, \bibinfo{person}{James Prather}, \bibinfo{person}{Andrew Luxton-Reilly}, \bibinfo{person}{Thezyrie Amarouche}, \bibinfo{person}{Brett~A Becker}, {and} \bibinfo{person}{Brent~N Reeves}.} \bibinfo{year}{2024}\natexlab{a}.
\newblock \showarticletitle{Prompt Problems: A new programming exercise for the generative AI era}. In \bibinfo{booktitle}{\emph{Proc. of the 55th ACM Technical Symp. on Computer Science Education V. 1}}.
\newblock


\bibitem[Denny et~al\mbox{.}(2024b)]%
        {denny2024computing}
\bibfield{author}{\bibinfo{person}{Paul Denny}, \bibinfo{person}{James Prather}, \bibinfo{person}{Brett~A. Becker}, \bibinfo{person}{James Finnie-Ansley}, \bibinfo{person}{Arto Hellas}, \bibinfo{person}{Juho Leinonen}, \bibinfo{person}{Andrew Luxton-Reilly}, \bibinfo{person}{Brent~N. Reeves}, \bibinfo{person}{Eddie~Antonio Santos}, {and} \bibinfo{person}{Sami Sarsa}.} \bibinfo{year}{2024}\natexlab{b}.
\newblock \showarticletitle{Computing Education in the Era of Generative AI}.
\newblock \bibinfo{journal}{\emph{Commun. ACM}} \bibinfo{volume}{67}, \bibinfo{number}{2} (\bibinfo{year}{2024}), \bibinfo{pages}{56–67}.
\newblock
\showISSN{0001-0782}


\bibitem[Dorodchi et~al\mbox{.}(2019)]%
        {dorodchi2019using}
\bibfield{author}{\bibinfo{person}{Mohsen Dorodchi}, \bibinfo{person}{Erfan Al-Hossami}, \bibinfo{person}{Aileen Benedict}, {and} \bibinfo{person}{Elise Demeter}.} \bibinfo{year}{2019}\natexlab{}.
\newblock \showarticletitle{Using synthetic data generators to promote open science in higher education learning analytics}. In \bibinfo{booktitle}{\emph{2019 IEEE Int. Conf. on Big Data}}. IEEE, \bibinfo{pages}{4672--4675}.
\newblock


\bibitem[Doughty et~al\mbox{.}(2024)]%
        {doughty2024comparative}
\bibfield{author}{\bibinfo{person}{Jacob Doughty}, \bibinfo{person}{Zipiao Wan}, \bibinfo{person}{Anishka Bompelli}, \bibinfo{person}{Jubahed Qayum}, \bibinfo{person}{Taozhi Wang}, \bibinfo{person}{Juran Zhang}, \bibinfo{person}{Yujia Zheng}, \bibinfo{person}{Aidan Doyle}, \bibinfo{person}{Pragnya Sridhar}, {et~al\mbox{.}}} \bibinfo{year}{2024}\natexlab{}.
\newblock \showarticletitle{A comparative study of AI-generated (GPT-4) and human-crafted MCQs in programming education}. In \bibinfo{booktitle}{\emph{Proc. of the 26th Australasian Computing Education Conf.}} \bibinfo{pages}{114--123}.
\newblock


\bibitem[Edwards et~al\mbox{.}(2023)]%
        {edwards2023review}
\bibfield{author}{\bibinfo{person}{John Edwards}, \bibinfo{person}{Kaden Hart}, \bibinfo{person}{Raj Shrestha}, {et~al\mbox{.}}} \bibinfo{year}{2023}\natexlab{}.
\newblock \showarticletitle{Review of CSEDM Data and Introduction of Two Public CS1 Keystroke Datasets}.
\newblock \bibinfo{journal}{\emph{J. of Educational Data Mining}} \bibinfo{volume}{15}, \bibinfo{number}{1} (\bibinfo{year}{2023}), \bibinfo{pages}{1--31}.
\newblock


\bibitem[Ettles et~al\mbox{.}(2018)]%
        {ettles2018common}
\bibfield{author}{\bibinfo{person}{Andrew Ettles}, \bibinfo{person}{Andrew Luxton-Reilly}, {and} \bibinfo{person}{Paul Denny}.} \bibinfo{year}{2018}\natexlab{}.
\newblock \showarticletitle{Common Logic Errors Made by Novice Programmers}. In \bibinfo{booktitle}{\emph{Proc. of the 20th Australasian Computing Education Conf.}} \bibinfo{publisher}{ACM}, \bibinfo{address}{New York, NY, USA}, \bibinfo{pages}{83–89}.
\newblock
\showISBNx{9781450363402}


\bibitem[Finnie-Ansley et~al\mbox{.}(2022)]%
        {Finnie-Ansley2022Robots}
\bibfield{author}{\bibinfo{person}{James Finnie-Ansley}, \bibinfo{person}{Paul Denny}, \bibinfo{person}{Brett~A. Becker}, \bibinfo{person}{Andrew Luxton-Reilly}, {and} \bibinfo{person}{James Prather}.} \bibinfo{year}{2022}\natexlab{}.
\newblock \showarticletitle{The Robots Are Coming: Exploring the Implications of OpenAI Codex on Introductory Programming}. In \bibinfo{booktitle}{\emph{Australasian Computing Education Conf.}} \bibinfo{publisher}{ACM}, \bibinfo{address}{New York, NY, USA}, \bibinfo{pages}{10–19}.
\newblock
\showISBNx{9781450396431}


\bibitem[Finnie-Ansley et~al\mbox{.}(2023)]%
        {finnie2023myai}
\bibfield{author}{\bibinfo{person}{James Finnie-Ansley}, \bibinfo{person}{Paul Denny}, \bibinfo{person}{Andrew Luxton-Reilly}, \bibinfo{person}{Eddie~Antonio Santos}, \bibinfo{person}{James Prather}, {and} \bibinfo{person}{Brett~A. Becker}.} \bibinfo{year}{2023}\natexlab{}.
\newblock \showarticletitle{My AI Wants to Know if This Will Be on the Exam: Testing OpenAI’s Codex on CS2 Programming Exercises}. In \bibinfo{booktitle}{\emph{Proc. of the 25th Australasian Computing Education Conf.}} \bibinfo{publisher}{ACM}, \bibinfo{pages}{97–104}.
\newblock
\showISBNx{9781450399418}


\bibitem[Flanagan et~al\mbox{.}(2022)]%
        {flanagan2022fine}
\bibfield{author}{\bibinfo{person}{Brendan Flanagan}, \bibinfo{person}{Rwitajit Majumdar}, {and} \bibinfo{person}{Hiroaki Ogata}.} \bibinfo{year}{2022}\natexlab{}.
\newblock \showarticletitle{Fine Grain Synthetic Educational Data: Challenges and Limitations of Collaborative Learning Analytics}.
\newblock \bibinfo{journal}{\emph{IEEE Access}}  \bibinfo{volume}{10} (\bibinfo{year}{2022}), \bibinfo{pages}{26230--26241}.
\newblock


\bibitem[Goel et~al\mbox{.}(2023)]%
        {goel2023llms}
\bibfield{author}{\bibinfo{person}{Akshay Goel}, \bibinfo{person}{Almog Gueta}, \bibinfo{person}{Omry Gilon}, \bibinfo{person}{Chang Liu}, \bibinfo{person}{Sofia Erell}, \bibinfo{person}{Lan~Huong Nguyen}, \bibinfo{person}{Xiaohong Hao}, \bibinfo{person}{Bolous Jaber}, \bibinfo{person}{Shashir Reddy}, \bibinfo{person}{Rupesh Kartha}, {et~al\mbox{.}}} \bibinfo{year}{2023}\natexlab{}.
\newblock \showarticletitle{Llms accelerate annotation for medical information extraction}. In \bibinfo{booktitle}{\emph{Machine Learning for Health (ML4H)}}. PMLR, \bibinfo{pages}{82--100}.
\newblock


\bibitem[Griffin(2019)]%
        {griffin2019designing}
\bibfield{author}{\bibinfo{person}{Jean~M. Griffin}.} \bibinfo{year}{2019}\natexlab{}.
\newblock \showarticletitle{Designing Intentional Bugs for Learning}. In \bibinfo{booktitle}{\emph{Proc. of the 2019 Conf. on United Kingdom \& Ireland Computing Education Research}}.
\newblock


\bibitem[Hoq et~al\mbox{.}(2024)]%
        {hoq2024detecting}
\bibfield{author}{\bibinfo{person}{Muntasir Hoq}, \bibinfo{person}{Yang Shi}, \bibinfo{person}{Juho Leinonen}, \bibinfo{person}{Damilola Babalola}, \bibinfo{person}{Collin Lynch}, \bibinfo{person}{Thomas Price}, {and} \bibinfo{person}{Bita Akram}.} \bibinfo{year}{2024}\natexlab{}.
\newblock \showarticletitle{Detecting ChatGPT-generated code submissions in a CS1 course using machine learning models}. In \bibinfo{booktitle}{\emph{Proceedings of the 55th ACM Technical Symposium on Computer Science Education V. 1}}. \bibinfo{pages}{526--532}.
\newblock


\bibitem[Hou et~al\mbox{.}(2024)]%
        {hou2023effects}
\bibfield{author}{\bibinfo{person}{Irene Hou}, \bibinfo{person}{Sophia Mettille}, \bibinfo{person}{Owen Man}, \bibinfo{person}{Zhuo Li}, \bibinfo{person}{Cynthia Zastudil}, {and} \bibinfo{person}{Stephen MacNeil}.} \bibinfo{year}{2024}\natexlab{}.
\newblock \showarticletitle{The Effects of Generative AI on Introductory Students’ Help-Seeking Preferences}. In \bibinfo{booktitle}{\emph{Australasian Computing Education Conference}}.
\newblock


\bibitem[Jordon et~al\mbox{.}(2022)]%
        {jordon2022synthetic}
\bibfield{author}{\bibinfo{person}{James Jordon}, \bibinfo{person}{Lukasz Szpruch}, \bibinfo{person}{Florimond Houssiau}, \bibinfo{person}{Mirko Bottarelli}, \bibinfo{person}{Giovanni Cherubin}, \bibinfo{person}{Carsten Maple}, \bibinfo{person}{Samuel~N Cohen}, {and} \bibinfo{person}{Adrian Weller}.} \bibinfo{year}{2022}\natexlab{}.
\newblock \showarticletitle{Synthetic Data--what, why and how?}
\newblock \bibinfo{journal}{\emph{arXiv preprint arXiv:2205.03257}} (\bibinfo{year}{2022}).
\newblock


\bibitem[Kazemitabaar et~al\mbox{.}(2024)]%
        {kazemitabaar2024codeaid}
\bibfield{author}{\bibinfo{person}{Majeed Kazemitabaar}, \bibinfo{person}{Runlong Ye}, \bibinfo{person}{Xiaoning Wang}, \bibinfo{person}{Austin~Zachary Henley}, \bibinfo{person}{Paul Denny}, \bibinfo{person}{Michelle Craig}, {and} \bibinfo{person}{Tovi Grossman}.} \bibinfo{year}{2024}\natexlab{}.
\newblock \showarticletitle{Codeaid: Evaluating a classroom deployment of an llm-based programming assistant that balances student and educator needs}. In \bibinfo{booktitle}{\emph{Proceedings of the CHI Conference on Human Factors in Computing Systems}}. \bibinfo{pages}{1--20}.
\newblock


\bibitem[Lau and Guo(2023)]%
        {lau2023from}
\bibfield{author}{\bibinfo{person}{Sam Lau} {and} \bibinfo{person}{Philip Guo}.} \bibinfo{year}{2023}\natexlab{}.
\newblock \showarticletitle{From ``Ban It Till We Understand It'' to ``Resistance is Futile'': How University Programming Instructors Plan to Adapt as More Students Use AI Code Generation and Explanation Tools such as ChatGPT and GitHub Copilot}. In \bibinfo{booktitle}{\emph{Proc. of the 2023 ACM Conf. on Int. Computing Education Research - Vol. 1}}. \bibinfo{publisher}{ACM}, \bibinfo{pages}{106–121}.
\newblock
\showISBNx{9781450399760}


\bibitem[Leinonen et~al\mbox{.}(2023)]%
        {leinonen2023comparing}
\bibfield{author}{\bibinfo{person}{Juho Leinonen}, \bibinfo{person}{Paul Denny}, \bibinfo{person}{Stephen MacNeil}, \bibinfo{person}{Sami Sarsa}, \bibinfo{person}{Seth Bernstein}, \bibinfo{person}{Joanne Kim}, \bibinfo{person}{Andrew Tran}, {and} \bibinfo{person}{Arto Hellas}.} \bibinfo{year}{2023}\natexlab{}.
\newblock \showarticletitle{Comparing Code Explanations Created by Students and Large Language Models}. In \bibinfo{booktitle}{\emph{Proc. of the 2023 Conf. on Innovation and Technology in Computer Science Education V. 1}}. \bibinfo{pages}{124–130}.
\newblock
\showISBNx{9798400701382}


\bibitem[Leinonen et~al\mbox{.}(2017)]%
        {leinonen2017preventing}
\bibfield{author}{\bibinfo{person}{Juho Leinonen}, \bibinfo{person}{Petri Ihantola}, {and} \bibinfo{person}{Arto Hellas}.} \bibinfo{year}{2017}\natexlab{}.
\newblock \showarticletitle{Preventing keystroke based identification in open data sets}. In \bibinfo{booktitle}{\emph{Proc. of the Fourth (2017) ACM Conference on Learning @ Scale}}. \bibinfo{pages}{101--109}.
\newblock


\bibitem[Li et~al\mbox{.}(2019)]%
        {li2019towards}
\bibfield{author}{\bibinfo{person}{Chen Li}, \bibinfo{person}{Emily Chan}, \bibinfo{person}{Paul Denny}, \bibinfo{person}{Andrew Luxton-Reilly}, {and} \bibinfo{person}{Ewan Tempero}.} \bibinfo{year}{2019}\natexlab{}.
\newblock \showarticletitle{Towards a Framework for Teaching Debugging}. In \bibinfo{booktitle}{\emph{Proc. of the Twenty-First Australasian Computing Education Conf.}} \bibinfo{pages}{79–86}.
\newblock


\bibitem[Li et~al\mbox{.}(2023)]%
        {li2023synthetic}
\bibfield{author}{\bibinfo{person}{Zhuoyan Li}, \bibinfo{person}{Hangxiao Zhu}, \bibinfo{person}{Zhuoran Lu}, {and} \bibinfo{person}{Ming Yin}.} \bibinfo{year}{2023}\natexlab{}.
\newblock \showarticletitle{Synthetic data generation with large language models for text classification: Potential and limitations}.
\newblock \bibinfo{journal}{\emph{arXiv preprint arXiv:2310.07849}} (\bibinfo{year}{2023}).
\newblock


\bibitem[Liffiton et~al\mbox{.}(2023)]%
        {liffiton2023codehelp}
\bibfield{author}{\bibinfo{person}{Mark Liffiton}, \bibinfo{person}{Brad~E Sheese}, \bibinfo{person}{Jaromir Savelka}, {and} \bibinfo{person}{Paul Denny}.} \bibinfo{year}{2023}\natexlab{}.
\newblock \showarticletitle{Codehelp: Using large language models with guardrails for scalable support in programming classes}. In \bibinfo{booktitle}{\emph{Proc. of the 23rd Koli Calling Int. Conf. on Computing Education Research}}.
\newblock


\bibitem[Liu et~al\mbox{.}(2024)]%
        {liu2024your}
\bibfield{author}{\bibinfo{person}{Jiawei Liu}, \bibinfo{person}{Chunqiu~Steven Xia}, \bibinfo{person}{Yuyao Wang}, {and} \bibinfo{person}{Lingming Zhang}.} \bibinfo{year}{2024}\natexlab{}.
\newblock \showarticletitle{Is Your Code Generated by ChatGPT Really Correct? Rigorous Evaluation of Large Language Models for Code Generation}.
\newblock \bibinfo{journal}{\emph{Advances in Neural Information Processing Systems}}  \bibinfo{volume}{36} (\bibinfo{year}{2024}).
\newblock


\bibitem[Long et~al\mbox{.}(2024)]%
        {long2024llms}
\bibfield{author}{\bibinfo{person}{Lin Long}, \bibinfo{person}{Rui Wang}, \bibinfo{person}{Ruixuan Xiao}, \bibinfo{person}{Junbo Zhao}, \bibinfo{person}{Xiao Ding}, \bibinfo{person}{Gang Chen}, {and} \bibinfo{person}{Haobo Wang}.} \bibinfo{year}{2024}\natexlab{}.
\newblock \showarticletitle{On LLMs-Driven Synthetic Data Generation, Curation, and Evaluation: A Survey}.
\newblock \bibinfo{journal}{\emph{arXiv preprint arXiv:2406.15126}} (\bibinfo{year}{2024}).
\newblock


\bibitem[MacNeil et~al\mbox{.}(2023)]%
        {macneil2023experiences}
\bibfield{author}{\bibinfo{person}{Stephen MacNeil}, \bibinfo{person}{Andrew Tran}, \bibinfo{person}{Arto Hellas}, \bibinfo{person}{Joanne Kim}, \bibinfo{person}{Sami Sarsa}, \bibinfo{person}{Paul Denny}, \bibinfo{person}{Seth Bernstein}, {and} \bibinfo{person}{Juho Leinonen}.} \bibinfo{year}{2023}\natexlab{}.
\newblock \showarticletitle{Experiences from Using Code Explanations Generated by Large Language Models in a Web Software Development E-Book}. In \bibinfo{booktitle}{\emph{Proc. of the ACM Technical Symp. on Computing Science Education}}. \bibinfo{publisher}{ACM}, \bibinfo{numpages}{6}~pages.
\newblock


\bibitem[MacNeil et~al\mbox{.}(2022)]%
        {macneil2022generating}
\bibfield{author}{\bibinfo{person}{Stephen MacNeil}, \bibinfo{person}{Andrew Tran}, \bibinfo{person}{Dan Mogil}, \bibinfo{person}{Seth Bernstein}, \bibinfo{person}{Erin Ross}, {and} \bibinfo{person}{Ziheng Huang}.} \bibinfo{year}{2022}\natexlab{}.
\newblock \showarticletitle{Generating Diverse Code Explanations Using the GPT-3 Large Language Model}. In \bibinfo{booktitle}{\emph{Proc. of the 2022 ACM Conf. on Int. Computing Education Research - Volume 2}}. \bibinfo{publisher}{ACM}, \bibinfo{pages}{37–39}.
\newblock
\showISBNx{9781450391955}


\bibitem[Markel et~al\mbox{.}(2023)]%
        {markel2023gpteach}
\bibfield{author}{\bibinfo{person}{Julia~M Markel}, \bibinfo{person}{Steven~G Opferman}, \bibinfo{person}{James~A Landay}, {and} \bibinfo{person}{Chris Piech}.} \bibinfo{year}{2023}\natexlab{}.
\newblock \showarticletitle{GPTeach: Interactive TA training with GPT-based students}. In \bibinfo{booktitle}{\emph{Proc. of the Tenth ACM Conf. on Learning @ Scale}}. \bibinfo{pages}{226--236}.
\newblock


\bibitem[McCauley et~al\mbox{.}(2008)]%
        {mccauley2008debugging}
\bibfield{author}{\bibinfo{person}{Ren{\'e}e McCauley}, \bibinfo{person}{Sue Fitzgerald}, \bibinfo{person}{Gary Lewandowski}, \bibinfo{person}{Laurie Murphy}, \bibinfo{person}{Beth Simon}, \bibinfo{person}{Lynda Thomas}, {and} \bibinfo{person}{Carol Zander}.} \bibinfo{year}{2008}\natexlab{}.
\newblock \showarticletitle{Debugging: a review of the literature from an educational perspective}.
\newblock \bibinfo{journal}{\emph{Computer Science Education}} \bibinfo{volume}{18}, \bibinfo{number}{2} (\bibinfo{year}{2008}), \bibinfo{pages}{67--92}.
\newblock


\bibitem[M{\o}ller et~al\mbox{.}(2024)]%
        {moller2024parrot}
\bibfield{author}{\bibinfo{person}{Anders~Giovanni M{\o}ller}, \bibinfo{person}{Arianna Pera}, \bibinfo{person}{Jacob Dalsgaard}, {and} \bibinfo{person}{Luca Aiello}.} \bibinfo{year}{2024}\natexlab{}.
\newblock \showarticletitle{The Parrot Dilemma: Human-Labeled vs. LLM-augmented Data in Classification Tasks}. In \bibinfo{booktitle}{\emph{Proceedings of the 18th Conference of the European Chapter of the Association for Computational Linguistics (Volume 2: Short Papers)}}. \bibinfo{pages}{179--192}.
\newblock


\bibitem[Moreno et~al\mbox{.}(2023)]%
        {moreno2023synthetic}
\bibfield{author}{\bibinfo{person}{Yaneth Moreno}, \bibinfo{person}{Anthony Montero}, \bibinfo{person}{Francisco Hidrobo}, {and} \bibinfo{person}{Saba Infante}.} \bibinfo{year}{2023}\natexlab{}.
\newblock \showarticletitle{Synthetic Data Generator for an E-Learning Platform in a Big Data Environment}. In \bibinfo{booktitle}{\emph{Int. Conf. in Information Technology and Education}}. Springer, \bibinfo{pages}{431--440}.
\newblock


\bibitem[Park et~al\mbox{.}(2024)]%
        {park2024chatlang}
\bibfield{author}{\bibinfo{person}{Jeiyoon Park}, \bibinfo{person}{Chanjun Park}, {and} \bibinfo{person}{Heuiseok Lim}.} \bibinfo{year}{2024}\natexlab{}.
\newblock \showarticletitle{ChatLang-8: An LLM-Based Synthetic Data Generation Framework for Grammatical Error Correction}.
\newblock \bibinfo{journal}{\emph{arXiv preprint arXiv:2406.03202}} (\bibinfo{year}{2024}).
\newblock


\bibitem[Perretta et~al\mbox{.}(2022)]%
        {perretta2022on}
\bibfield{author}{\bibinfo{person}{James Perretta}, \bibinfo{person}{Andrew DeOrio}, \bibinfo{person}{Arjun Guha}, {and} \bibinfo{person}{Jonathan Bell}.} \bibinfo{year}{2022}\natexlab{}.
\newblock \showarticletitle{On the use of mutation analysis for evaluating student test suite quality}. In \bibinfo{booktitle}{\emph{Proc. of the 31st ACM SIGSOFT Int. Symp. on Software Testing and Analysis}}. \bibinfo{pages}{263–275}.
\newblock


\bibitem[Piech et~al\mbox{.}(2015)]%
        {piech2015deep}
\bibfield{author}{\bibinfo{person}{Chris Piech}, \bibinfo{person}{Jonathan Bassen}, \bibinfo{person}{Jonathan Huang}, \bibinfo{person}{Surya Ganguli}, \bibinfo{person}{Mehran Sahami}, \bibinfo{person}{Leonidas~J Guibas}, {and} \bibinfo{person}{Jascha Sohl-Dickstein}.} \bibinfo{year}{2015}\natexlab{}.
\newblock \showarticletitle{Deep knowledge tracing}.
\newblock \bibinfo{journal}{\emph{Advances in neural information processing systems}}  \bibinfo{volume}{28} (\bibinfo{year}{2015}).
\newblock


\bibitem[Prather et~al\mbox{.}(2023)]%
        {prather2023robots}
\bibfield{author}{\bibinfo{person}{James Prather}, \bibinfo{person}{Paul Denny}, \bibinfo{person}{Juho Leinonen}, \bibinfo{person}{Brett~A. Becker}, \bibinfo{person}{Ibrahim Albluwi}, \bibinfo{person}{Michelle Craig}, \bibinfo{person}{Hieke Keuning}, \bibinfo{person}{Natalie Kiesler}, \bibinfo{person}{Tobias Kohn}, \bibinfo{person}{Andrew Luxton-Reilly}, \bibinfo{person}{Stephen MacNeil}, \bibinfo{person}{Andrew Petersen}, \bibinfo{person}{Raymond Pettit}, \bibinfo{person}{Brent~N. Reeves}, {and} \bibinfo{person}{Jaromir Savelka}.} \bibinfo{year}{2023}\natexlab{}.
\newblock \showarticletitle{The Robots Are Here: Navigating the Generative AI Revolution in Computing Education}. In \bibinfo{booktitle}{\emph{Proc. of the 2023 Working Group Reports on Innovation and Technology in Computer Science Education}}. \bibinfo{publisher}{ACM}, \bibinfo{pages}{108–159}.
\newblock
\showISBNx{9798400704055}


\bibitem[Prather et~al\mbox{.}(2024)]%
        {prather2024widening}
\bibfield{author}{\bibinfo{person}{James Prather}, \bibinfo{person}{Brent Reeves}, \bibinfo{person}{Juho Leinonen}, \bibinfo{person}{Stephen MacNeil}, \bibinfo{person}{Arisoa~S Randrianasolo}, \bibinfo{person}{Brett Becker}, \bibinfo{person}{Bailey Kimmel}, \bibinfo{person}{Jared Wright}, {and} \bibinfo{person}{Ben Briggs}.} \bibinfo{year}{2024}\natexlab{}.
\newblock \showarticletitle{The Widening Gap: The Benefits and Harms of Generative AI for Novice Programmers}.
\newblock \bibinfo{journal}{\emph{arXiv preprint arXiv:2405.17739}} (\bibinfo{year}{2024}).
\newblock


\bibitem[Sarsa et~al\mbox{.}(2022a)]%
        {sarsa2022automatic}
\bibfield{author}{\bibinfo{person}{Sami Sarsa}, \bibinfo{person}{Paul Denny}, \bibinfo{person}{Arto Hellas}, {and} \bibinfo{person}{Juho Leinonen}.} \bibinfo{year}{2022}\natexlab{a}.
\newblock \showarticletitle{Automatic Generation of Programming Exercises and Code Explanations Using Large Language Models}. In \bibinfo{booktitle}{\emph{Proc. of the 2022 ACM Conf. on Int. Computing Education Research - Volume 1}}. \bibinfo{publisher}{ACM}, \bibinfo{pages}{27–43}.
\newblock
\showISBNx{9781450391948}


\bibitem[Sarsa et~al\mbox{.}(2022b)]%
        {sarsa2022empirical}
\bibfield{author}{\bibinfo{person}{Sami Sarsa}, \bibinfo{person}{Juho Leinonen}, \bibinfo{person}{Arto Hellas}, {et~al\mbox{.}}} \bibinfo{year}{2022}\natexlab{b}.
\newblock \showarticletitle{Empirical Evaluation of Deep Learning Models for Knowledge Tracing: Of Hyperparameters and Metrics on Performance and Replicability}.
\newblock \bibinfo{journal}{\emph{Journal of Educational Data Mining}} \bibinfo{volume}{14}, \bibinfo{number}{2} (\bibinfo{year}{2022}).
\newblock


\bibitem[Tang et~al\mbox{.}(2023)]%
        {tang2023does}
\bibfield{author}{\bibinfo{person}{Ruixiang Tang}, \bibinfo{person}{Xiaotian Han}, \bibinfo{person}{Xiaoqian Jiang}, {and} \bibinfo{person}{Xia Hu}.} \bibinfo{year}{2023}\natexlab{}.
\newblock \showarticletitle{Does synthetic data generation of llms help clinical text mining?}
\newblock \bibinfo{journal}{\emph{arXiv preprint arXiv:2303.04360}} (\bibinfo{year}{2023}).
\newblock


\bibitem[Tran et~al\mbox{.}(2023)]%
        {tran2023generating}
\bibfield{author}{\bibinfo{person}{Andrew Tran}, \bibinfo{person}{Kenneth Angelikas}, \bibinfo{person}{Egi Rama}, \bibinfo{person}{Chiku Okechukwu}, \bibinfo{person}{David~H Smith}, {and} \bibinfo{person}{Stephen MacNeil}.} \bibinfo{year}{2023}\natexlab{}.
\newblock \showarticletitle{Generating multiple choice questions for computing courses using large language models}. In \bibinfo{booktitle}{\emph{2023 IEEE Frontiers in Education Conference (FIE)}}. IEEE, \bibinfo{pages}{1--8}.
\newblock


\bibitem[Wagner et~al\mbox{.}(2024)]%
        {wagner2024power}
\bibfield{author}{\bibinfo{person}{Stefan~Sylvius Wagner}, \bibinfo{person}{Maike Behrendt}, \bibinfo{person}{Marc Ziegele}, {and} \bibinfo{person}{Stefan Harmeling}.} \bibinfo{year}{2024}\natexlab{}.
\newblock \showarticletitle{The Power of LLM-Generated Synthetic Data for Stance Detection in Online Political Discussions}.
\newblock \bibinfo{journal}{\emph{arXiv preprint arXiv:2406.12480}} (\bibinfo{year}{2024}).
\newblock


\bibitem[Whalley et~al\mbox{.}(2021)]%
        {whalley2021novice}
\bibfield{author}{\bibinfo{person}{Jacqueline Whalley}, \bibinfo{person}{Amber Settle}, {and} \bibinfo{person}{Andrew Luxton-Reilly}.} \bibinfo{year}{2021}\natexlab{}.
\newblock \showarticletitle{Novice Reflections on Debugging}. In \bibinfo{booktitle}{\emph{Proc. of the 52nd ACM Technical Symp. on Computer Science Education}}. \bibinfo{publisher}{ACM}, \bibinfo{address}{New York, NY, USA}, \bibinfo{pages}{73–79}.
\newblock
\showISBNx{9781450380621}


\bibitem[Zastudil et~al\mbox{.}(2023)]%
        {zastudil2023generative}
\bibfield{author}{\bibinfo{person}{Cynthia Zastudil}, \bibinfo{person}{Magdalena Rogalska}, \bibinfo{person}{Christine Kapp}, \bibinfo{person}{Jennifer Vaughn}, {and} \bibinfo{person}{Stephen MacNeil}.} \bibinfo{year}{2023}\natexlab{}.
\newblock \showarticletitle{Generative ai in computing education: Perspectives of students and instructors}. In \bibinfo{booktitle}{\emph{IEEE Frontiers in Education Conference}}. IEEE, \bibinfo{pages}{1--9}.
\newblock


\bibitem[Zhan et~al\mbox{.}(2023)]%
        {zhan2023synthetic}
\bibfield{author}{\bibinfo{person}{Chen Zhan}, \bibinfo{person}{Oscar~Blessed Deho}, \bibinfo{person}{Xuwei Zhang}, \bibinfo{person}{Srecko Joksimovic}, {and} \bibinfo{person}{Maarten de Laat}.} \bibinfo{year}{2023}\natexlab{}.
\newblock \showarticletitle{Synthetic data generator for student data serving learning analytics: A comparative study}.
\newblock \bibinfo{journal}{\emph{Learning Letters}} (\bibinfo{year}{2023}).
\newblock


\end{thebibliography}


\end{document}